EURADOS Working Group 6, Computational Dosimetry, a history of promoting good practice via intercomparisons and training

Rick Tanner[1], Stefano Agosteo[2] and Hans Rabus[3]

1 United Kingdom Health Security Agency, Chilton, Didcot, Oxon OX11 0RQ, United Kingdom

2 Politecnico di Milano, Milano, Italy

3 Physikalisch Technische Bundesanstalt (PTB), Braunschweig and Berlin, Germany

EURADOS was founded in 1982 (Rühm et al., 2020) and since that time it has managed research programmes in radiation protection via a set of Working Groups covering the topics of primary interest in the field. The fields covered by the working groups have evolved over the years, with new working groups being formed as new areas of interest are identified, with working groups being discontinued as their usefulness declines. Ever present in EURADOS has been a Computational Dosimetry working group that has brought together scientists working in numerical methods.

Whilst EURADOS Working Group 6 (WG6) is strongly associated with Monte Carlo methods, its remit has always covered other numerical methods used in radiation protection, especially unfolding methods for neutron spectrometry (Alevra et al., 1990). The radiation modelling methods covered have also included deterministic calculations, though those are not commonly used in radiation protection nowadays because of the availability of powerful computers and Monte Carlo codes that are widely distributed. Over the past four decades the membership of WG6 has evolved with too many contributors to list. In this time, it has, however, only been chaired by Siegfried Wagner, Bernd Siebert, Gianfranco Gualdrini, Rick Tanner and Hans Rabus.

During the lifetime of WG6, the field of computational dosimetry has evolved dramatically. When WG6 started, the members of the working group were largely writing their own codes to perform Monte Carlo or deterministic calculations, and most of the computational dosimetry experts working in radiation protection in Europe were concentrated within the working group. The codes used had the great merit of independence, since each one was a unique attempt to solve a problem. Comparisons between such codes were hence necessary as a quality assurance check. Poor application of the code, however, was not such a problem, because only the originator knew how to use it.

One of the earliest tasks of the computational dosimetry was an intercomparison of unfolding codes (Alevra et al., 1990) which promoted high quality neutron spectrum unfolding at a number of European laboratories. In that case, the laboratories involved in the intercomparison were the ones that were developing the methods for neutron spectrometry, but more recently it has been necessary to perform an open intercomparison (Gómez-Ros et al., 2018, 2022), because the methods are used widely across many dosimetry laboratories. The results were largely good, but in several cases, they were clearly incorrect in ways that are hard to explain.

Though not technically an intercomparison, WG6's highest profile achievement was the generation of the data for the operational dose quantities and the protection quantities as described in ICRU Report 51 (ICRU, 1987) and ICRP Publication 60 (ICRP, 1991). This yielded the joint ICRU and ICRP publications ICRU Report 57 (ICRU, 1998) and ICRP Publication 74 (ICRP, 1996). The operational quantity conversion coefficients in those publications remain in use today, though the protection quantity conversion coefficients have been superseded by those in ICRP Publication 116 (Petoussi-Henss et al., 2010). The generation of the conversion coefficients was, in essence, an intercomparison between the members of WG6, since at that time these calculations were at the limit of what could be achieved with the codes and computers that were available: participants presented their results and refined methods to get good agreement, and a "reference solution" produced using a weighted sum of the results. There was not even standardization using reference phantoms at that time. This concept of a "reference solution" has become one of the key issues in computational intercomparisons: what is the "right" answer? At that time, coupled calculations that included the transport of the secondary electrons to produce true estimates of the absorbed dose was in its infancy. It is hence fascinating to see where we are now; with recent calculations of using full secondary particle transport, voxel and now mesh phantoms to represent the individual, and wide ranges of particle type and energy. However, it must be remembered that the calculations performed by EURADOS WG6 to produce ICRU Report 57 and ICRP Publication 74 were cutting edge at the time, and that EURADOS co-ordinated work produced data that are still in use today, almost 30 years later.

Codes produced by an individual scientists suffer from the limited effort available for code development, and the tendency of those codes to retire from the field with their originator. Those codes also tended not to cover all types of radiation and hence could not be applied for all problems. Scientists writing their own codes from scratch are now largely replaced by large teams producing codes that can be applied to a wide range of problems, with relatively little expertise on behalf of the user. The main codes in use in radiation protection today are the product of many person-years of coding to produce a programme that covers a huge range of applications, with a potentially user-friendly interface. Additionally, running Monte Carlo codes used to require mainframe computers, whereas today the codes can be run on PCs or laptops, though for the most cpu intensive calculations PC clusters are often preferred. In the past, running enough starting particles to get convergence was a problem, but now there are more likely to be issues of exceeding the random number stride and getting correlated histories.

In the early days of EURADOS, numerical methods were largely restricted to the activities of WG6. However, today all WGs have a computational element to their work programme. Running Monte Carlo simulations is a basic part of the skillset of most young scientists in radiation protection and dosimetry and a high fraction of scientific papers in radiation protection and dosimetry have a Monte Carlo component. This leads to the important question: how reliable are the results?

The results in Monte Carlo are limited by the available data for cross-sections and materials, and the physics models implemented in them. They are also limited by the available computing power. It is, however, clear that, in the hands of experts, a range of codes and scientists can produce consistent results even for difficult computational problems (Petoussi-Henss et al., 2010). But how reliable are the codes in the hands of less experience and less expert users?

These questions led to a series of computational intercomparisons that were run by EURADOS WG6. These sought to test how well the users of computational codes in radiation protection and dosimetry actually use their codes. This special issue of Radiation Measurements summarizes the recent intercomparisons performed with the remit and in collaboration with other EURADOS Working Groups. However, even though performing intercomparisons has become a major part of the EURADOS work programme, as recognized in our Strategic Research Agendas (Rühm et al., 2016, Harrison et al., 2021), EURADOS intercomparisons on computational methods have been running throughout the history of this WG. In particular, the QUADOS (**QU**ality **A**ssurance of computational tools for **DOS**imetry) set of eight problems formed a concerted set of intercomparisons (Tanner et al., 2004, Siebert et al., 2006), that became the basis for much of the future work of WG6. These culminated in a 2003 workshop in Bologna that drew together experts and young scientists working in radiation protection and dosimetry to discuss the often impressive, and frequently concerning accuracy of the submitted solutions.

The papers gathered together in this special issue of Radiation Measurements carry this work programme forward. They cover more complex problems in terms of geometry, particle types, energy ranges, coupled calculations and also scale, with the possibility of performing Monte Carlo calculations on micro and nano dosimetric scales now feasible. They also required computer power that was not feasible during the QUADOS intercomparison.

A summary of the exercises is provided in the last article of the Special Issue (Rabus et al., 2022), which presents the findings and common conclusions from the ten articles reporting the results of each exercise (De Saint-Hubert et al., 2021, 2022; Eakins et al., 2021; Gómez-Ros et al., 2021, 2022; Huet et al., 2022; Rabus et al., 2021; Villagrasa et al., 2022; Zankl et al., 2021b, 2021c). One of these issues was the correct assessment of bone marrow dose, which prompted the inclusion of an article in this special issue explaining the ICRP-recommended method for bone marrow dosimetry (Zankl et al., 2021a).

Whilst WG6 has in many respects led the way for EURADOS in terms of intercomparisons, one area in which we are left behind is accreditation. The many intercomparisons that EURADOS has run for personal dosemeters now constitute evidence for ISO 17025 (Petrovic et al., 2020) accreditation of those services. Perhaps this is a future direction of travel for Monte Carlo calculations? Accredited Monte Carlo or Computational Dosimetry expert?

Acknowledgement

We would like to thank the three previous chairs of EURADOS Working Group 6, Siegfried Wagner, Bernd Siebert and Gianfranco Gualdrini for developing the work programme for Computational Dosimetry within EURADOS and helping to build the Working Group to what it is now. Without their efforts computational dosimetry would not have become so integral to EURADOS' activities.